\documentclass[final,3p,twocolumn,sort&compress]{elsarticle}

\usepackage{amsmath}
\usepackage{amssymb}
\usepackage[figurename=Fig.,labelsep=period]{caption}
\usepackage[colorlinks,citecolor=blue,urlcolor=blue,linkcolor=blue]{hyperref}
\usepackage{slashed}

\begin{document}

\begin{frontmatter}



\title{A modified dyon solution in a non-Abelian gauge model}

\author[tusur]{A.Yu.~Loginov}
\ead{a.yu.loginov@tusur.ru}

\address[tusur]{Laboratory of Applied Mathematics and Theoretical Physics, Tomsk State
                University of Control Systems and Radioelectronics, 634050 Tomsk, Russia}

\begin{abstract}
A modified $SU(2)$ Georgi-Glashow  model  is  considered  here, and it is shown
that besides the well-known  Julia-Zee dyon solution, the model can also have a
modified dyon solution.
The properties of the modified  dyon  solution are studied using analytical and
numerical methods.
A comparative analysis of the modified  dyon  and the Julia-Zee dyon shows that
their properties are significantly different.
In particular, except for the BPS case, the energy and electric  charge  of the
modified dyons exceed considerably those of the Julia-Zee dyons.
At the same time, the energy and  electric  charge  of  the  modified  dyon are
bounded for all admissible  parameter  values,  whereas  those of the Julia-Zee
dyon can be arbitrarily large in the BPS case.
\end{abstract}


\end{frontmatter}

\section{Introduction}                                            \label{seq:I}

Among the various soliton solutions of field theory, there  are those that have
an electric charge.
These    soliton   solutions  can    exist    in    both  $(2 + 1)$-dimensional
\cite{paul, khare_rao_227,  hong,  jw1,  jw2,  bazeia_1991, ptz_plb_339, ghosh,
arth_tchr_prd_54, deshaies_2006, lgn_2014,navarro_2017} and $(3+1)$-dimensional
\cite{jz_1975, prs_1975, cpns_1975, bkt_1999, rduchr_2006, klee, lee_yoon_1991,
ardoz_2009, tamaki_2014,brihaye_prd_2014,gulamov_2014,gulamov_2015,lshnir_2019,
loginov_2020} gauge models.
To ensure the finiteness of the energy of two-dimensional electrically  charged
solitons, $(2 + 1)$-dimensional gauge models must include the Chern-Simons term
\cite{JT,schonfeld, DJT}, whereas the Maxwell term may be missing.
Due to the Chern-Simons  term, the gauge field in these models is topologically
massive, and therefore the electric field of the solitons is short-range.
In contrast, the electric field  of  three-dimensional  solitons is long-range,
since there is no Chern-Simons term in $(3 + 1)$ dimensions and the gauge field
models include only the Maxwell term.

The electrically charged solitons  can be divided into two classes: topological
solitons     \cite{jz_1975, prs_1975, cpns_1975, bkt_1999, rduchr_2006}     and
nontopological    solitons    \cite{klee, lee_yoon_1991,ardoz_2009,tamaki_2014,
brihaye_prd_2014,gulamov_2014,gulamov_2015, lshnir_2019, loginov_2020}.
The properties of the topological and  nontopological solitons are considerably
different.
Because of topological triviality, the existence of  nontopological solitons is
entirely due to dynamical factors.
In particular, a conserved Noether charge must exist in the corresponding field
model, and the self-interaction potential  of  the scalar fields has to be of a
special form \cite{coleman, lee}.
For a  fixed  value  of  the  Noether  charge,  the  field  configuration  of a
nontopological  soliton  is  a   stationary  point  of  the  energy  functional
\cite{fried, fried1, fried2}.
As a consequence, the  energy  and  the  Noether  charge  of the nontopological
soliton are related  by  a  differential relation, which determines a number of
the soliton's properties.

In  contrast,  topological  solitons   have   topologically   nontrivial  field
configurations, which prevents them from transitioning into lower energy states
\cite{Manton}.
Hence,  the  presence  of  a  potential  term  and   Noether  charge  is  not a
necessary  condition  for   the   existence   of  three-dimensional topological
solitons, as  can  be  seen in the example of the BPS monopole \cite{prs_1975}.
The  best  known  example  of   an   electrically   charged   three-dimensional
topological soliton  is   the  dyon  solution  \cite{jz_1975}   in  the $SU(2)$
Georgi-Glashow model \cite{GG_1972}.
Similar to nontopological solitons,  a  dyon  solution  is  a  stationary point
(minimum) of the energy functional for a fixed value of the Noether  (electric)
charge.
It follows that the the energy  and  the  electric  charge of the dyon are also
related by a  differential  relation  \cite{lg_plb_2021},  as  is  the case for
the nontopological  solitons.
As a result, except for the  BPS  case,  the  energy and electric charge of the
dyon cannot be arbitrarily large \cite{bkt_1999}.

In \cite{lg_plb_2018, lg_plb_2018b},  a  soliton  system  consisting  of vortex
and Q-ball  components interacting via an electric field is described.
Being a source of an  electric  field,  a  dyon  can also interact with another
charged field, as a result of which a new soliton system can arise.
In this paper, we consider a  soliton  system (modified dyon) arising  from the
interaction   of   the   Julia-Zee   dyon   \cite{jz_1975}   and  an  isospinor
self-interacting scalar field.
The modified dyon consists  of a core (deformed Julia-Zee dyon) surrounded by a
cloud of  a  charged scalar field, and its properties differ significantly from
those of the Julia-Zee dyon.

Throughout this paper, we use the natural units $c = 1$, $\hbar = 1$.

\section{Lagrangian and field equations of the model}            \label{seq:II}

We consider  a  modified  Georgi-Glashow model with a non-Abelian  gauge group
$SU(2)$.
The model contains two types of self-interacting scalar fields, one of which is
transformed according to the adjoint representation of the gauge group $SU(2)$,
and    the    other    according    to    the    fundamental    representation.
The Lagrangian of the model takes the form
\begin{flalign}
\mathcal{L} &=-\frac{1}{4}F_{\mu \nu}^{a}F^{a\,\mu \nu}+\frac{1}{2}
\left(D_{\mu }\phi ^{a}\right) \left( D^{\mu }\phi ^{a}\right)
\nonumber \\
&+\left( D_{\mu }\Phi \right) ^{\dag }\left( D^{\mu}\Phi \right)-V\left(
\phi^{a}\phi^{a}\right)-U\left(\Phi^{\dag}\Phi\right),             \label{II:1}
\end{flalign}
where
\begin{equation}
F_{\mu \nu}^{a}=\partial_{\mu}A_{\nu}^{a}-\partial_{\nu}A_{\mu}^{a}
              -g\epsilon^{abc}A_{\mu}^{b}A_{\nu}^{c}               \label{II:2}
\end{equation}
is the non-Abelian field strength,
\begin{equation}
D_{\mu}\phi^{a}=\partial_{\mu}\phi^{a}-g\epsilon^{abc}A_{\mu}^{b}\phi^{c}
                                                                   \label{II:3}
\end{equation}
is the covariant derivative of the  isovector scalar field $\boldsymbol{\phi}$,
and
\begin{equation}
D_{\mu}\Phi=\partial_{\mu}\Phi + i g A_{\mu}^{a}\frac{\tau_{a}}{2}\Phi
                                                                   \label{II:4}
\end{equation}
is the covariant derivative of the isospinor scalar field $\Phi$.
The  self-interaction  potentials  of the isovector and isospinor scalar fields
are
\begin{equation}
V\left(\phi^{a}\phi^{a}\right) = \frac{\lambda}{4}\left(\phi^{a}
\phi^{a}-v^{2}\right)^{2}                                          \label{II:5}
\end{equation}
and
\begin{equation}
U\left(\Phi^{\dag}\Phi\right) = M^{2}\Phi^{\dag}\Phi-\frac{G}{2}\left(
\Phi^{\dag}\Phi\right)^{2}+\frac{H}{3}\left(\Phi^{\dag}\Phi\right)^{3},
                                                                   \label{II:6}
\end{equation}
respectively, where $M$ is  a  mass  parameter, and $\lambda$, $G$, and $H$ are
coupling constants.
Note that the  six-order  potential  in  Eq.~(\ref{II:6}) is of the form widely
used in models having Q-ball solutions.

Eq.~(\ref{II:5}) tells us that the  potential  $V$ reaches a zero minimum value
at non-zero $\left\vert\boldsymbol{\phi} \right\vert = v$.
At the same time, we assume  that  the  potential  $U$  has a global minimum at
$\Phi = 0$.
For this to hold, the  parameters of the potential $U$ in Eq.~(\ref{II:6}) must
satisfy the inequality $3 G^{2} < 16 H M^{2}$.
The nonzero vacuum value  of  the  isovector  scalar  field $\boldsymbol{\phi}$
leads to a spontaneous  violation  of  the  gauge  group  $SU(2)$ to an Abelian
(electromagnetic) subgroup $U(1)$.
As a result of the  Higgs  mechanism,  the  spectrum  of  small fluctuations of
fields relative to the  vacuum contains one neutral  massless vector field, two
charged massive vector fields with mass $m_{V} = gv$, one neutral massive Higgs
field with mass $m_{H} = \sqrt{2 \lambda} v$,  and  two  charged massive scalar
fields with  mass $m_{\Phi} = M$.

Using standard methods of field theory, we  find  the field equations of model:
(\ref{II:1})
\begin{flalign}
& D_{\mu}F^{a\,\mu \nu} = J^{a\,\nu},                             \label{II:7a}
 \\
& D_{\mu}D^{\mu}\phi^{a} = -\lambda \left(\phi^{b}\phi^{b}-v^{2}\right)
\phi ^{a},                                                        \label{II:7b}
 \\
& D_{\mu}D^{\mu}\Phi  = -\left[ M^{2}-G\left(\Phi^{\dag}\Phi\right)
+H\left(\Phi^{\dag}\Phi \right)^{2}\right] \Phi,                  \label{II:7c}
\end{flalign}
where the covariantly conserved current
\begin{flalign}
J^{a\,\nu } &= g \epsilon_{abc}\phi_{b}D^{\nu}\phi_{c} + i 2^{-1} g
\left[\Phi^{\dag}\tau^{a}D^{\nu}\Phi \right.
\nonumber  \\
&\bigl. -\left( D^{\nu }\Phi \right) ^{\dag }\tau^{a}\Phi \bigr].  \label{II:8}
\end{flalign}
Later on, we shall also need the  expression  for the energy-momentum tensor of
the model,
\begin{flalign}
T_{\mu \nu } & = -F_{\mu \rho }^{a}F_{\nu }^{a\,\rho }+\left( D_{\mu }\phi
^{a}\right) \left( D_{\nu }\phi ^{a}\right)
\nonumber  \\
& +D_{\mu }\Phi ^{\dag }D_{\nu }\Phi +D_{\nu }\Phi ^{\dag }D_{\mu }\Phi
\nonumber  \\
& +\eta _{\mu \nu }\left[ \frac{1}{4}F_{\rho \tau }^{a}F^{a\,\rho \tau }-%
\frac{1}{2}\left( D_{\rho }\phi ^{a}\right) \left( D^{\rho }\phi ^{a}\right)
\right.
\nonumber  \\
& \biggl. -D_{\mu }\Phi ^{\dag }D^{\mu }\Phi +V\left( \phi ^{a\ast }\phi
^{a}\right) +U\left( \Phi ^{\dag }\Phi \right) \biggr],            \label{II:9}
\end{flalign}
where the metric tensor $\eta_{\mu \nu} = \text{diag}(1, -1, -1, -1)$.

\section{Some properties of the modified dyon solution}         \label{seq:III}

From Eqs.~(\ref{II:5})  and  (\ref{II:9}),  it  follows  that to possess finite
energy, a field configuration of model (\ref{II:1}) must satisfy the asymptotic
condition $\underset{r \rightarrow  \infty} {\lim} \left\vert \boldsymbol{\phi}
\right\vert = v$.
From a topological point of view, this condition is equivalent to the existence
of a mapping of the infinitely distant space  sphere  $S^{2}_{\infty}$  to  the
vacuum sphere $S_{\text{vac}}^{2}\!: \left\vert \boldsymbol{\phi} \right\vert =
v$.
Since the  second  homotopy  group  $\pi_{2}\left( S^{2} \right) = \mathbb{Z}$,
the  mapping $S^{2}_{\infty}  \rightarrow S^{2}_{\text{vac}}$  is topologically
nontrivial   and   splits  into   different   topological   classes,  which are
characterised  by the integer winding number $n$.

In the absence of the  isospinor scalar  field $\Phi$, model (\ref{II:1}) turns
into the $SU(2)$ Georgi–Glashow model.
It is well known that in the topological sector with the winding number $n= 1$,
the $SU(2)$ Georgi–Glashow model  has  two  topological  soliton solutions: the
't~Hooft–Polyakov monopole \cite{hooft_74, polyakov_74} and the Julia--Zee dyon
\cite{jz_1975}.
Both the 't~Hooft--Polyakov  monopole  and  the  Julia--Zee  dyon have the same
magnetic charge $Q_{M}$, but  the  dyon  also  possesses  an electrical charge.
The electric charge of the dyon  can lie in the range from zero to some maximum
value, which tends to infinity in the BPS case.

The  inclusion  of   the   self-interacting   isospinor   scalar   field $\Phi$
modifies  the   Georgi–Glashow   model   leading    to    model   (\ref{II:1}).
Our aim is to ascertain  the possibility of the existence of a dyon solution in
model (\ref{II:1}) and to investigate the properties of this solution.
To find a dyon solution, we use the following ansatz:
\begin{subequations}                                              \label{III:1}
\begin{flalign}
A^{a 0}& = n^{a} v j\left(r\right),                             \label{III:1a}
  \\
A^{a i}& = \epsilon^{a i m} n^{m}\frac{1-u\left(r\right)}{gr},  \label{III:1b}
  \\
\phi ^{a} &= n^{a}v h \left(r\right),                            \label{III:1c}
  \\
\Phi  &= v f\left(r\right)\left(i n^{a}\tau_{a}\right)\Phi _{0}, \label{III:1d}
\end{flalign}
\end{subequations}
where the unit vector $n^{a}= x^{a}/r$, and $\Phi_{0}$  is a constant isospinor
normalized to unity: $\Phi_{0}^{\dagger}\Phi _{0} = 1$.
The ansatz (\ref{III:1}) corresponds to the  most general spherically symmetric
(modulo a global gauge transformation)  field  configuration  that  is odd with
respect to the inversion $\mathbf{x} \rightarrow -\mathbf{x}$.
Note   that    Eqs.~(\ref{III:1a}) -- (\ref{III:1c})    describe    the   field
configuration  of   the   Julia--Zee   dyon,   while   Eqs.~(\ref{III:1b})  and
(\ref{III:1d}) coincide  in  form  with  the  ansatz for the sphaleron solution
 \cite{klmn_1984} in the Standard Model.

We now  introduce  the  dimensionless  radial  variable $\rho = m_{V} r$, where
$m_{V} = g v$ is the mass of the electrically charged gauge bosons.
Substituting       ansatz       (\ref{III:1})     into     field      equations
(\ref{II:7a})--(\ref{II:7c}),  we  obtain  a  system  of nonlinear differential
equations for the ansatz functions:
\begin{flalign}
& j^{\prime \prime }\left(\rho\right)+\frac{2}{\rho}j^{\prime}\left(\rho
\right)
- \left( \frac{1}{2}f\left( \rho \right) ^{2}+\frac{2}{\rho ^{2}}u(\rho
)^{2}\right) j(\rho )=0,                                         \label{III:2a}
\end{flalign}
\begin{flalign}
& u^{\prime \prime }\left( \rho \right) -\frac{u(\rho )\left( u(\rho
)^{2}-1\right) }{\rho ^{2}}-\bigl( 2^{-1}f\left( \rho \right) ^{2}  \bigr.
 \nonumber \\
& \bigl. + h(\rho )^{2}-j(\rho )^{2} \bigr) u(\rho
)-2^{-1}f\left( \rho \right) ^{2}=0,                             \label{III:2b}
\end{flalign}
\begin{flalign}
& h^{\prime \prime }\left( \rho \right) +\frac{2}{\rho }h^{\prime }\left( \rho
\right) -\frac{2}{\rho ^{2}}u(\rho )^{2}h(\rho )
 \nonumber \\ &
+\kappa \left( 1-h(\rho )^{2}\right) h(\rho )=0,               \label{III:2c}
\end{flalign}
\begin{flalign}
& f^{\prime \prime }\left( \rho \right) +\frac{2}{\rho }f^{\prime }\left(
\rho \right) -\frac{\left( 1+u(\rho )\right) ^{2}}{2\rho ^{2}}f\left( \rho
\right)                                                          \label{III:2d}
 \\
& -\!\left[ \left( \tilde{M}^{2}\!-\!\tilde{G}f(\rho)^{2}\!+\!\tilde{H}
f(\rho)^{4}\right)\!-\!2^{-2}j(\rho)^{2}\right]f\left(\rho\right)\!=\!0,
                                                                    \nonumber
\end{flalign}
where the dimensionless parameters are  $\kappa  = \lambda/g^{2}$, $\tilde{M} =
M/m_{V}$, $\tilde{G} = G/g^{2}$, and $\tilde{H} = H v^{2}/g^{2}$.
We next obtain the expression for the energy $E = \int T_{00}d^{3}x$ of a field
configuration in terms of the ansatz functions:
\begin{flalign}
& E=m_{M}\int\limits_{0}^{\infty }\left[ \frac{u^{\prime }(\rho )^{2}}{
\rho ^{2}}+\frac{1}{2}j^{\prime }(\rho )^{2}+\frac{1}{2}h^{\prime }(\rho
)^{2}+f^{\prime }(\rho )^{2}\right.
\nonumber \\
&\left. +\frac{\left(1-u(\rho )^{2}\right)^{2}}{2\rho^{4}}
 +\frac{\left( j(\rho )^{2}+h(\rho )^{2}\right) }{\rho ^{2}}
 u(\rho)^{2}\right.
\nonumber \\
&\left. +\frac{\left( 1+u(\rho )\right) ^{2}}{2\rho ^{2}}f(\rho )^{2}+\frac{
1}{4}j(\rho )^{2}f(\rho )^{2}+\frac{\kappa }{4}\left( 1-h(\rho )^{2}\right)
^{2}\right.
\nonumber \\
&\biggl. +\tilde{M}^{2}f(\rho )^{2} - \frac{\tilde{G}}{2}f(\rho )^{4} +
\frac{\tilde{H}}{3}f(\rho)^{6}\biggr]\rho^{2}d\rho,                 \label{III:3}
\end{flalign}
where $m_{M} = 4 \pi v g^{-1}$ is the mass of the BPS monopole \cite{prs_1975}.

The dyon solution must  be  regular  at  the  origin  and also must have finite
energy.
These two conditions together with Eqs.~(\ref{III:1}) and (\ref{III:3}) lead us
to the boundary conditions for the ansatz functions:
\begin{subequations}                                              \label{III:4}
\begin{align}
&j(0)=0,\qquad \underset{\rho\rightarrow \infty}{\lim}j(\rho)=j_{\infty},
                                                                 \label{III:4a}
 \\
&u(0)=1,\qquad \underset{\rho\rightarrow \infty}{\lim}u(\rho)=0, \label{III:4b}
 \\
&h(0)=0,\qquad \underset{\rho\rightarrow \infty}{\lim}h(\rho)=1, \label{III:4c}
 \\
&f(0)=0,\qquad \underset{\rho\rightarrow \infty}{\lim}f(\rho)=0, \label{III:4d}
\end{align}
\end{subequations}
where $j_{\infty}$ is a finite value.

For large  $\rho$, Eqs.~(\ref{III:2a}) -- (\ref{III:2d})  are linearized, which
makes it possible to study the asymptotics of the dyon solution.
As a  result,  we  conclude  that  in  Eq.~(\ref{III:4a}),  the  limiting value
$j_{\infty}$ must satisfy the condition
\begin{equation}
\left\vert j_{\infty}\right\vert <\min \bigl(2\tilde{M},1\bigr).  \label{III:5}
\end{equation}
Otherwise, the asymptotics for  large $\rho$  of  at  least  one  of the ansatz
functions  $u(\rho)$  and  $f(\rho)$   will  be  oscillatory,  resulting  in an
infinite energy of the field configuration.

Eq.~(\ref{III:1c}) tells  us  that  the   Higgs   field  $\boldsymbol{\phi}$ is
invariant under gauge transformations  of the $U(1)$ subgroup corresponding  to
rotations around  the unit vector $n^{a} = x^{a}/r$ in the isospace.
As a result, the dyon  solution  possesses a long-range gauge (electromagnetic)
field with the field strength  tensor $F_{\mu \nu } = v^{-1}\phi ^{a}F_{\mu \nu
}^{a}$.
The intensities of the electric and  magnetic  fields  of the dyon solution are
expressed in terms of the ansatz functions as
\begin{equation}
E_{i}=v^{-1}F_{0i}^{a}\phi ^{a}=-gv^{2}h\left( \rho \right)
j^{\prime}\left( \rho \right) n_{i}                               \label{III:6}
\end{equation}
and
\begin{flalign}
B_{i} & = -\left( 2v\right) ^{-1}\epsilon _{ijk}F_{jk}^{a}\phi ^{a}
\nonumber \\
&= -gv^{2}\frac{h\left( \rho \right) }{\rho ^{2}}\left( 1-u\left( \rho
\right) ^{2}\right) n_{i},                                        \label{III:7}
\end{flalign}
respectively.
Using Eqs.~(\ref{III:6}) and (\ref{III:6}), we  obtain the  expressions for the
electric and magnetic charges of the dyon solution:
\begin{flalign}
& Q_{E} = \oint_{S_{\infty }^{2}}d^{2}S_{i}E_{i}=-\frac{4\pi }{g}
\underset{\rho \rightarrow \infty}{\lim}
\rho^{2} j^{\prime}\left(\rho\right)                              \label{III:8}
\end{flalign}
and
\begin{equation}
Q_{M}=\oint_{S_{\infty }^{2}} d^{2}S_{i}B_{i} = -\frac{4\pi}{g},  \label{III:9}
\end{equation}
where    the   negative   magnetic   charge   for   the   field   configuration
with the positive winding number $n = 1$  is  due  to  the  choice  of the sign
before the gauge coupling  constant  $g$  in  Eqs.(\ref{II:2}) and (\ref{II:3})
which is similar to that used in \cite{Manton, gdrol_1978}.
From Eq.~(\ref{III:8}), we  then  obtain  the  large  $\rho$ asymptotics of the
ansatz function $j(\rho)$ in terms of the electric charge of the dyon
\begin{equation}
j\left(\rho\right) \sim j_{\infty}+\frac{g}{4\pi}\frac{Q_{E}}{\rho}.
                                                                 \label{III:10}
\end{equation}

We can write the dyon's energy (\ref{III:3}) as the sum of five terms
\begin{equation}
E = E^{\left( E\right)}+E^{\left( T\right)}+E^{\left(B\right)}
+E^{\left(G\right)}+E^{\left(P\right)},                          \label{III:11}
\end{equation}
where
\begin{equation}
E^{\left( E\right) }=m_{M}\int\limits_{0}^{\infty }\left[ \frac{1}{2}
j^{\prime }\left( \rho \right) ^{2}+\frac{j\left( \rho \right) ^{2}u\left(
\rho \right) ^{2}}{\rho ^{2}}\right] \rho ^{2}d\rho              \label{III:12}
\end{equation}
is the energy of the electric field,
\begin{equation}
E^{\left( T\right) }=m_{M}\int\limits_{0}^{\infty }\left[ \frac{1}{4}
j\left(\rho\right)^{2}f\left(\rho\right)^{2}\right]\rho^{2}d\rho \label{III:13}
\end{equation}
is the kinetic energy of the isospinor scalar field $\Phi$,
\begin{equation}
E^{\left( B\right) }=m_{M}\int\limits_{0}^{\infty }\left[ \frac{u^{\prime
}\left( \rho \right) ^{2}}{\rho ^{2}}+\frac{\bigl( 1-u\left( \rho \right)
^{2}\bigr)^{2}}{2\rho ^{4}}\right] \rho ^{2}d\rho                \label{III:14}
\end{equation}
is the energy of the magnetic field,
\begin{flalign}
& E^{\left( G\right) } = m_{M}\int\limits_{0}^{\infty }\left[ \frac{1}{2}
h^{\prime }\left( \rho \right) ^{2}+f\left( \rho \right) ^{\prime 2}\right.
  \nonumber \\
&\biggl. +\frac{u\left( \rho \right) ^{2}h\left( \rho \right) ^{2}}{\rho ^{2}
}+\frac{\left( 1+u\left( \rho \right) \right) ^{2}}{2\rho ^{2}}f\left( \rho
\right) ^{2}\biggr] \rho ^{2}d\rho                               \label{III:15}
\end{flalign}
is the gradient part of the energy, and
\begin{flalign}
& E^{\left(P\right)} = m_{M}\int\limits_{0}^{\infty }\biggl[\frac{\kappa}{4}
\bigl( 1-h\left( \rho \right) ^{2}\bigr)^{2}+\tilde{M}
^{2}f\left( \rho \right) ^{2}\biggr.
  \nonumber  \\
& \biggl. -\frac{\tilde{G}}{2}f\left( \rho \right) ^{4}+\frac{\tilde{H}}{3}
f\left( \rho \right) ^{6}\biggr] \rho ^{2}d\rho                  \label{III:16}
\end{flalign}
is the potential part of the energy.
Using  Gauss's   law    (\ref{III:2a}),    we   then   can   express   the  sum
$E^{\left( E \right)} + E^{\left(T \right)}$ in terms of the electric charge of
the dyon
\begin{equation}
E^{\left(E\right)}+E^{\left(T\right)}=-2^{-1}vj_{\infty }Q_{E}.  \label{III:17}
\end{equation}

Similar to the energy in Eq.~(\ref{III:11}), the Lagrangian $L=\int \mathcal{L}
d^{3}x$  can   also   be     written    as     a    linear    combination    of
Eqs.~(\ref{III:12}) -- (\ref{III:16})
\begin{equation}
L = E^{\left( E\right) }+E^{\left( T\right) }-E^{\left( B\right)}
-E^{\left(G\right) }-E^{\left( P\right)}.                        \label{III:18}
\end{equation}
Since the  Lagrangian  density  (\ref{II:1})  does not depend on time for field
configurations  (\ref{III:1}),  any  solution   of    system  (\ref{III:2a}) --
(\ref{III:2d}) is   a   stationary   point   of  the Lagrangian~(\ref{III:18}).
Let $j(\rho)$, $u(\rho)$, $h(\rho)$, and  $f(\rho)$  be  a  solution  of system
(\ref{III:2a}) -- (\ref{III:2d})   satisfying   the   boundary   conditions  in
Eq.~(\ref{III:4}).
After the scale transformation $\rho\rightarrow \varkappa \rho$ of the argument
of the solution, the Lagrangian  $L$  becomes a function of the scale parameter
$\varkappa$.
From the above, it follows  that  the function $L (\varkappa)$ has a stationary
point at $\varkappa = 1$,  and  hence  its  derivative vanishes  at this point:
$\left. dL/d\varkappa\right\vert_{\varkappa = 1} = 0$.

We can easily find  the  transformation  laws  of  the constituent parts of the
dyon's energy under the rescaling $\rho \rightarrow\varkappa \rho$: $E^{\left(E
\right) } \rightarrow \varkappa^{-1} E^{\left( E\right)}$, $E^{\left( T\right)}
\rightarrow \varkappa^{-3}E^{\left(T\right)}$,  $E^{\left(B\right)} \rightarrow
\varkappa E^{\left( B\right)}$, $E^{\left( G \right)}\rightarrow \varkappa^{-1}
E^{\left(G\right)}$, and $E^{\left(P\right)}\rightarrow\varkappa ^{-3}E^{\left(
P\right)}$.
Using Eq.~(\ref{III:18}) and these  transformation  laws,  we obtain the virial
relation for the dyon solution
\begin{equation}
E^{\left( E\right) }+E^{\left( B\right) }+3E^{\left( T\right) }-
E^{\left(G\right)}-3E^{\left(P\right)}=0.                        \label{III:19}
\end{equation}

Eqs.~(\ref{III:11}),  (\ref{III:17}),  and   (\ref{III:18})  tell  us  that the
Lagrangian, the  energy, and  the  electric charge of the dyon are connected by
the relation
\begin{equation}
-L = E + v j_{\infty} Q_{E}.                                     \label{III:20}
\end{equation}
We know that the dyon solution is  an unconditional extremum (stationary point)
of the Lagrangian $L$.
Then, Eq.~(\ref{III:20}) tells us that the  dyon solution is also a conditional
extremum of the energy functional $E$  at  a fixed value of the electric charge
$Q_{E}$,  with  $v j_{\infty}$  being  the  Lagrange multiplier.
We  define  the  Noether   charge  of  the  dyon  as  $Q_{N}  =  g^{-1} Q_{E}$.
Varying Eq.~(\ref{III:20}) and equating  the  variation  to zero, we obtain the
important differential relation
\begin{equation}
dE/dQ_{N} = g\,dE/dQ_{E} = \Omega,                               \label{III:21}
\end{equation}
where
\begin{equation}
\Omega =-gvj_{\infty }=-m_{V}j_{\infty }.                        \label{III:22}
\end{equation}
Eqs.~(\ref{III:21}) and (\ref{III:22})  tell  us  that  the  derivative  of the
dyon's energy with respect to its Noether (electric) charge  is proportional to
the limiting value of the ansatz function $j(\rho)$ at infinity.
Note  that   a   differential   relation   similar  to  Eq.~(\ref{III:21}) is a
characteristic property  of  nontopological   solitons   and, in particular, of
Q-balls \cite{coleman}.

The system of differential equations  (\ref{III:2a}) -- (\ref{III:2d}) together
with boundary conditions (\ref{III:4}) is  a  boundary  value  problem   on the
semi-infinite interval $\rho\in\left[0, \infty\right)$.
This  problem   admits   the   trivial  solution $f(\rho)  =  0$;  in this case
the solution to  the  remaining  Eqs.~(\ref{III:2a}) -- (\ref{III:2c})  of  the
system  coincides  with  the  Julia-Zee  dyon  solution \cite{jz_1975}  of  the
$SU(2)$ Georgi-Glashow model \cite{GG_1972}.
The ansatz function $f(\rho)$ enters Eqs.~(\ref{III:2a}) -- (\ref{III:2c}) only
via the quadratic combination $f(\rho)^{2}$.
We can therefore suppose that if $f(\rho)$ is small enough, its backreaction on
Eqs.~(\ref{III:2a}) -- (\ref{III:2c}) is all  the  smaller, and the solution to
the  boundary  value  problem  will also exist for non-zero $f(\rho)$.

A qualitative analysis of the boundary value problem reveals that the amplitude
of $f(\rho)$ is roughly determined  by  the position $f_{\min}$ of the non-zero
minimum  of the potential $U\left(f\right)=\tilde{M}^{2}f^{2} - 2^{-1}\tilde{G}
f^{4} + 3^{-1}\tilde{H} f^{6}$.
In addition, $U(f_{\min})$  must   be  sufficiently  close  to  zero, otherwise
$f(\rho)$ increases indefinitely as $\rho \rightarrow \infty$.
These  two  conditions  allow  us   to  obtain  approximate  estimates  for the
parameters $\tilde{G}$ and $\tilde{H}$:    $\tilde{G} \lessapprox 4\tilde{M}^{2
} f_{\min}^{-2}$  and  $\tilde{H}  \lessapprox  3 \tilde{M}^{2} f_{\min}^{-4}$.
In addition, the smallness condition  for  $f(\rho)$ can be written as $f_{\min
}^{2} \ll 1$, since for the Julia-Zee dyon, the  maximum values  of  the ansatz
functions $u(\rho)$ and $h(\rho)$ are equal to unity.

\section{Numerical results}
                                  \label{seq:IV}
In this section, we  present  numerical  results  concerning the modified  dyon
solution of model (\ref{II:1}).
To obtain them, we  use  the  numerical  methods  provided  in the {\sc{Maple}}
package \cite{maple}.
To check the correctness of the  numerical results, we use Eqs.~(\ref{III:17}),
(\ref{III:19}), and (\ref{III:21}).
It follows  from  Eqs.~(\ref{III:2a}) -- (\ref{III:2d}) and  (\ref{III:4}) that
the  dimensionless  ansatz   functions  $u(\rho)$,  $j(\rho)$,  $h(\rho)$,  and
$f(\rho)$ depend  on  the five dimensionless parameters: $\kappa$, $\tilde{M}$,
$\tilde{G}$, $\tilde{H}$, and $\tilde{\Omega} = -j_{\infty}$.
Furthermore, from  Eqs.~(\ref{III:3})  and  (\ref{III:8}) it  follows  that the
dimensionless combinations $\tilde{Q}_{E} = g Q_{E}$ (rescaled electric charge)
and $\tilde{E} = g^{2} m_{V}^{-1} E = g v^{-1} E$ (rescaled energy) also depend
only on these five parameters.
The numerical results presented  here correspond to the parameters $\tilde{M} =
0.3$, $\tilde{G} = 12.5$, and $\tilde{H} = 326$.
These parameters  correspond to self-interaction  potential (\ref{II:6}) having
an almost zero positive minimum at sufficiently small $f \approx 0.17$.
In accordance with the above, this property of the potential (\ref{II:6}) makes
possible the existence of dyon solutions in model (\ref{II:1}).
Note that for $v = 240\,\text{GeV}$  and  $g = 1/3$, which are typical Standard
Model values corresponding to $m_{V}= 80 \,\text{GeV}$, we obtain the following
values of the dimensional parameters:  $M = 24\,  \text{GeV}$,  $G = 1.39$, and
$H = 6.3\times 10^{-4}\, \text{GeV}^{-2}$.

\begin{figure}[tbp]
\includegraphics[width=7.8cm]{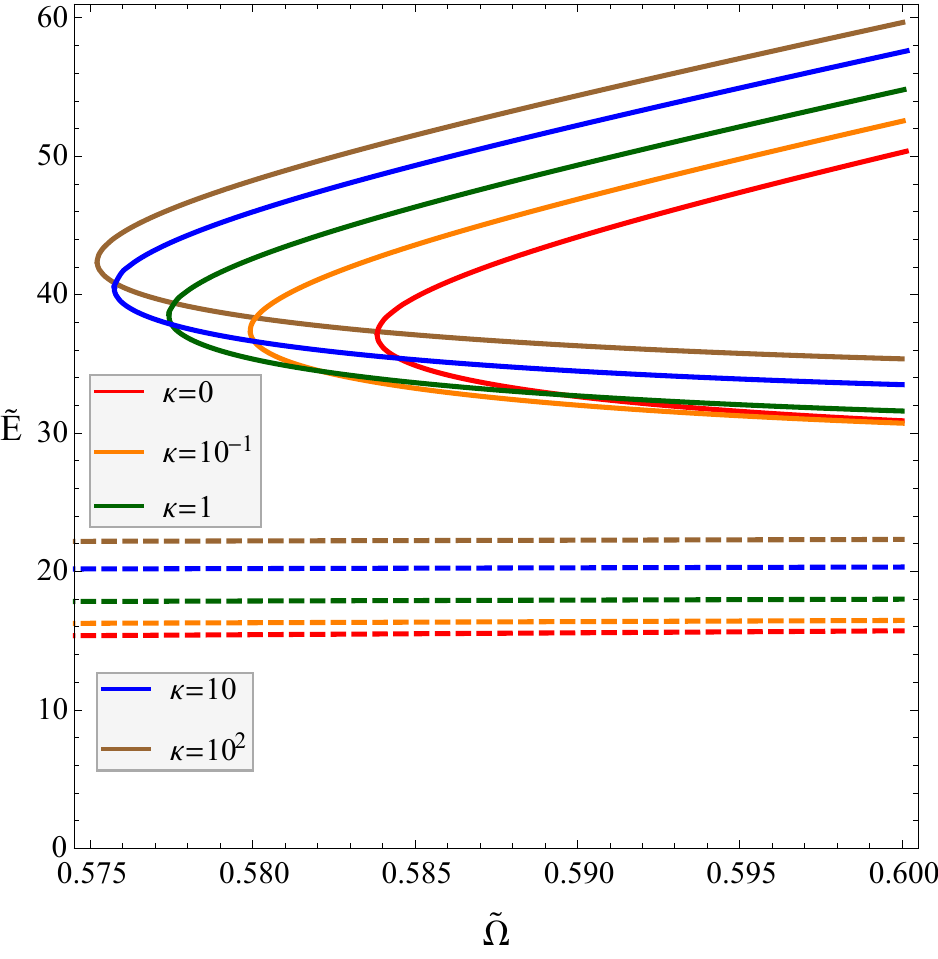}
\caption{\label{fig1}       Dependence  of  the  rescaled  energy  $\tilde{E} =
g^{2} m_{V}^{-1} E$  on  the   parameter  $\tilde{\Omega} = m_{V}^{-1} \Omega =
-j_{\infty}$  for different values of  the parameter $\kappa = g^{-2} \lambda$.
The solid  lines  correspond  to  the  modified  dyons,  and  the  dashed lines
correspond to the Julia-Zee dyons.}
\end{figure}

Figure~\ref{fig1} presents the dependence  of  the  energy  $\tilde{E}$ for the
modified dyon solution of model (\ref{II:1}) on the parameter  $\tilde{\Omega}$
for different values of the parameter $\kappa = g^{-2} \lambda$.
In addition, Fig.~\ref{fig1} shows similar dependencies for the Julia-Zee dyon.
Firstly, we see that the  modified dyon solution  exists only in a narrow range
$\bigl[ \tilde{\Omega}_{\min}, 2 \tilde{M} \bigr]$  of  values of the parameter
$\tilde{\Omega}$.
The right boundary of  the  interval  $\bigl[\tilde{\Omega}_{\min}, 2 \tilde{M}
\bigr]$ is  in  agreement  with  Eq.~(\ref{III:5}),  while  the  left  boundary
decreases monotonically  with  an  increase  in  $\kappa$ and tends to a finite
limit as $\kappa \rightarrow \infty$.
The width of this interval is quite  small and is  only $0.02 - 0.025$, whereas
the Julia-Zee  dyon  exists in a much wider wide  interval  $\left[0, 1\right]$
of $|\tilde{\Omega}|$.

It follows from Fig.~\ref{fig1} that the $\tilde{E}(\tilde{\Omega})$ curves for
the modified and Julia-Zee dyons behave completely differently.
Indeed, we see  that  for  the  modified  dyon, the $\tilde{E}(\tilde{\Omega})$
curves consist of two (lower and upper) branches that join at $\tilde{\Omega} =
\tilde{\Omega}_{\min}$.
For a given  $\kappa$, the  energy  of  the  modified  dyon  reaches  a minimum
(maximum)  value  on  the  lower   (upper)   branch  at  the  maximum  value of
$\tilde{\Omega} = 2 \tilde{M}$.
Such a behaviour of  the $\tilde{E}(\tilde{\Omega})$ curves is similar to their
behaviour for an  electrically  charged  Q-ball  and  completely different from
that for the Julia-Zee dyon.
Note   that   in   Fig.~\ref{fig1},   the   $\tilde{E}(\tilde{\Omega})$  curves
corresponding to  the Julia-Zee  dyon  are  practically  indistinguishable from
straight lines.
This  is  because  in   Fig.~\ref{fig1},   the   width   of   the   interval of
$\tilde{\Omega}$ is much less than unity.

\begin{figure}[tbp]
\includegraphics[width=7.8cm]{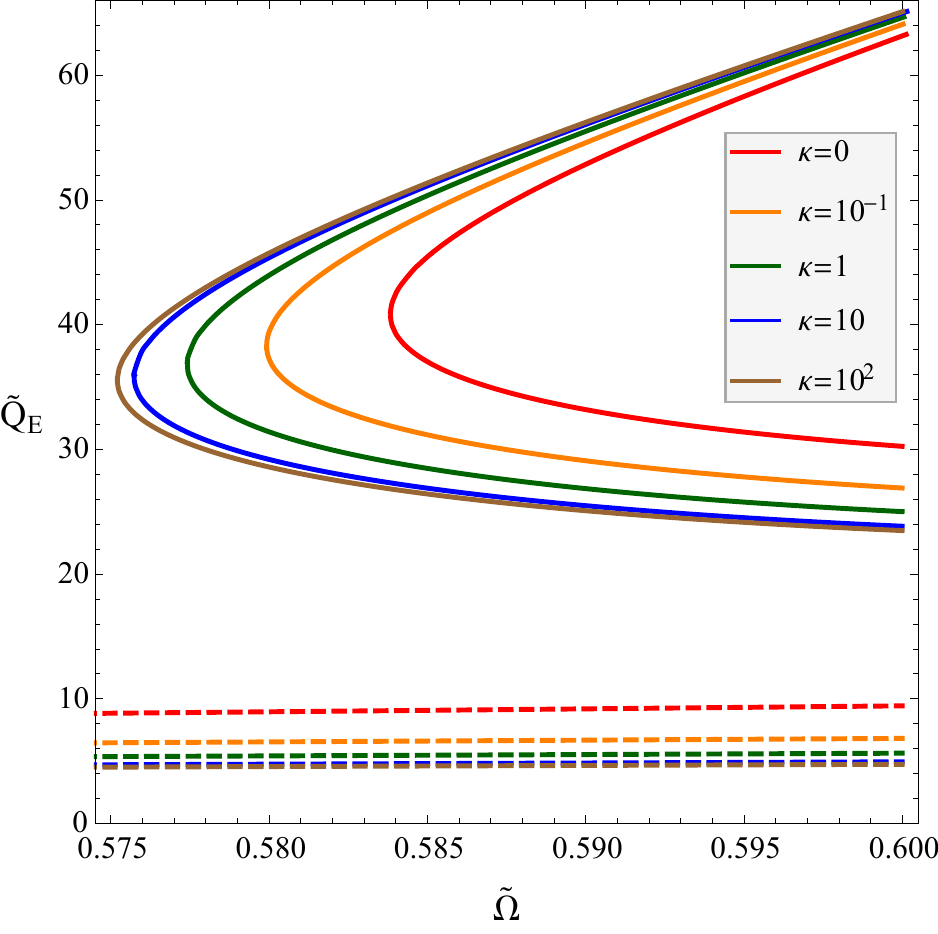}
\caption{\label{fig2}   Dependence of the rescaled electric charge $\tilde{Q}_{
E} = g Q_{E}$ on the parameter $\tilde{\Omega} = m_{V}^{-1}\Omega =-j_{\infty}$
for different   values   of   the   parameter   $\kappa   =   g^{-2}  \lambda$.
The solid  lines  correspond  to  the  modified  dyons,  and  the  dashed lines
correspond to the Julia-Zee dyons.}
\end{figure}

Figure~\ref{fig2} presents  the dependence of the electric charge $\tilde{Q}_{E
}$ of the modified dyon solution  on  $\tilde{\Omega}$  for different values of
$\kappa$.
We  see  that the  behavior  of  the  $\tilde{Q}_{E}(\tilde{\Omega})$ curves is
similar to that of the $\tilde{E}(\tilde{\Omega})$ curves for both the modified
and Julia-Zee dyon solutions.
The difference is that for the modified  dyon  solution, the $\tilde{E}(\tilde{
\Omega})$ curves intersect each other, whereas the $\tilde{Q}_{E}(\tilde{\Omega
})$ curves do not.
As a result, for the modified  dyon  solution, the minimum and maximum energies
are reached at $\tilde{\Omega} = 2\tilde{M}$, and both of them increase with an
increase in $\kappa$.
But the  minimum  (maximum)  electric  charge  of  the  modified  dyon solution
decreases (increases) with an increase in $\kappa$.

\begin{figure}[tbp]
\includegraphics[width=7.8cm]{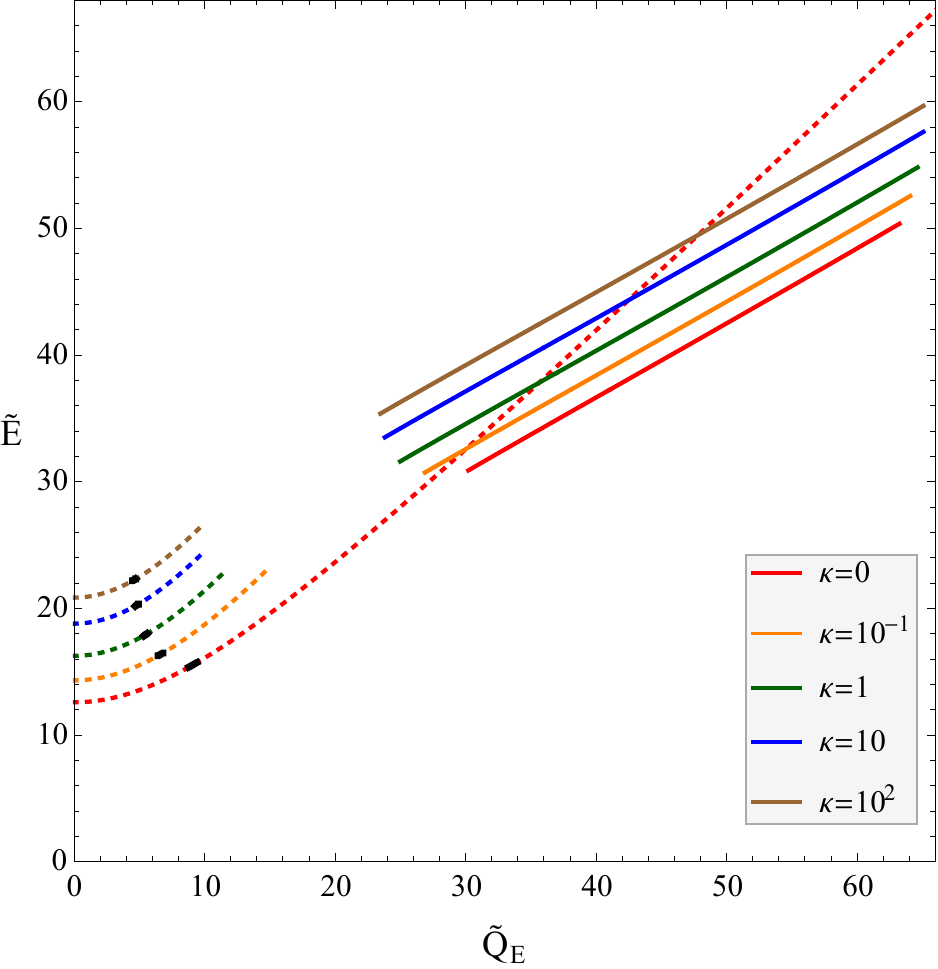}
\caption{\label{fig3} The $\tilde{E}(\tilde{Q}_{E})$  dependences for different
$\kappa$.    The solid lines  correspond  to the modified dyons, and the dotted
lines correspond to the Julia-Zee dyons.    The  black  segments  on the dotted
curves  correspond  to  the  ranges  of the Julia-Zee  dyon solutions presented
in Figs.~\ref{fig1} and \ref{fig2}.}
\end{figure}

Figure~\ref{fig3}  shows  the  dependences  of  the energy  $\tilde{E}$  on the
electric charge $\tilde{Q}_{E}$  for  both  the  modified and  Julia-Zee dyons.
Note that in accordance   with   \cite{bkt_1999},  the  energy  $\tilde{E}$ and
the electric charge $\tilde{Q}_{E}$ of the Julia-Zee dyon  can take arbitrarily
large values only when $\kappa = 0$ (the BPS case), whereas  there  are maximum
possible values  for  both  $\tilde{E}$  and  $\tilde{Q}_{E}$  when $\kappa>0$.
We see that the $\tilde{E}(\tilde{Q}_{E})$  curves  are substantially different
for the modified and Julia-Zee dyons.
Firstly, except for the case $\kappa = 0$, the energies and electric charges of
the   modified  dyon  solutions  exceed  the  maximum  possible  values  of the
corresponding Julia-Zee dyon solutions.
Secondly, for the modified dyon solutions, the curves $\tilde{E}(\tilde{Q}_{E})$
are visually indistinguishable  from  straight lines,  which is not the case for
the Julia-Zee dyon solutions.
This  is  because   in   Figs.~\ref{fig1}   and   \ref{fig2},  the  interval of
variation  of  $\tilde{\Omega}$  is  much  less  than $\tilde{\Omega}$  itself:
$\Delta \tilde{\Omega}/\tilde{\Omega} \sim 1/30$, and the above property of the
solid curves in Fig.~\ref{fig3}  follows  from  Eq.~(\ref{III:21}) rewritten as
\begin{equation}
d\tilde{E}/d\tilde{Q}_{E} = \tilde{\Omega}.                        \label{IV:1}
\end{equation}
Thirdly, unlike  the  Julia-Zee dyon,  the  energy  and  electric charge of the
modified dyon cannot be arbitrarily large even when the parameter $\kappa = 0$.

Figure~\ref{fig3} allows us to address the subject of stability of the modified
dyon solution.
Consider the  family  of  the  half-lines $a + b \tilde{Q}_{E}$ coming from the
points of the dotted curve that correspond to a given value of $\kappa$.
It is understood that the  parameter $a$  depends  on  the  origin point of the
half-line.
Then, it follows from Fig.~\ref{fig3} that for all the points $p$ of the dotted
curves, the half-lines $a(p) + \tilde{Q}_{E}$ pass over the corresponding solid
lines.
Hence, the  modified  dyon   is  stable  to  the  transition into the Julia-Zee
dyon accompanied by emission of the charged vector bosons of mass $m_{V}= g v$.
In contrast, for all the  points  $p$  of  the  dotted  curves corresponding to
$\kappa \ge 0.1$, the half-lines $a(p)+2\tilde{M} \tilde{Q}_{E}$ pass under the
corresponding solid lines, where the appearance of the factor $2$ is due to the
fact that the electric charges of the components of the scalar isodublet $\Phi$
are $\pm g/2$.
The same is true for points $p$ of the dotted curve corresponding to $\kappa=0$
provided that $\tilde{Q}_{E}\left(p\right) \lesssim 25$.
It follows that  the  modified  dyon  is  unstable  to  the transition into the
Julia-Zee dyon accompanied  by  emission  of  the charged scalar bosons of mass
$M$.

\begin{figure}[tbp]
\includegraphics[width=7.8cm]{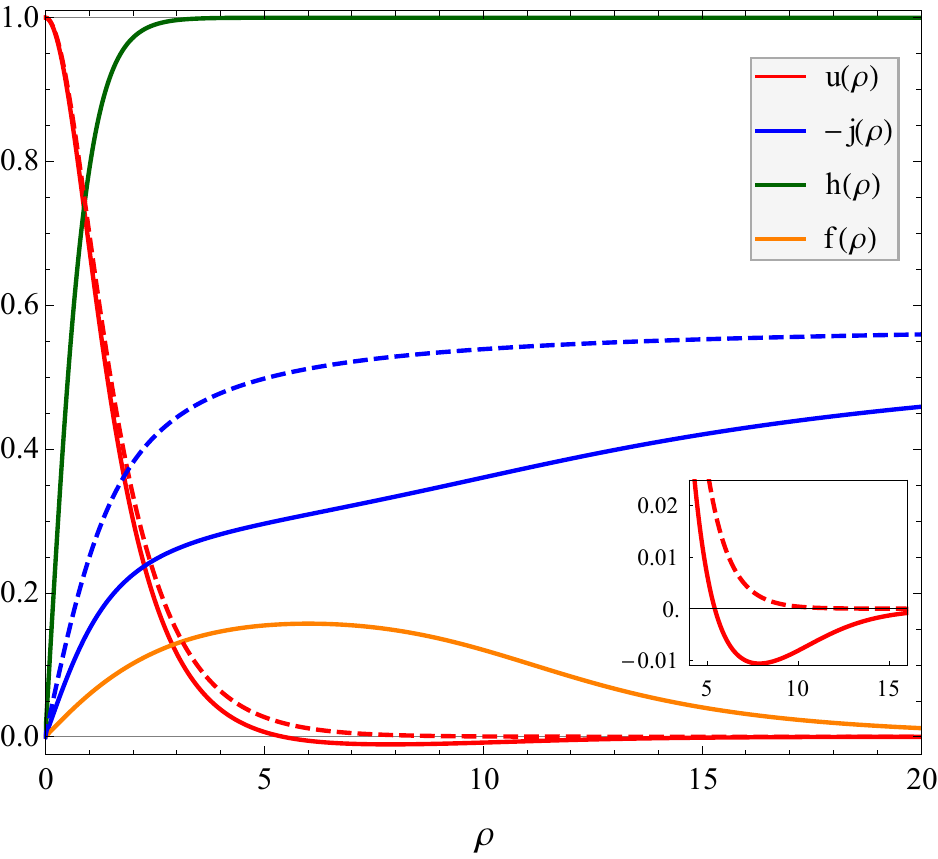}
\caption{\label{fig4}   Ansatz functions of the dyon solutions corresponding to
the parameters $\tilde{\Omega}=-j_{\infty}=0.58$ and $\kappa=g^{-2} \lambda=2$.
The solid lines correspond to the  modified dyon solution, and the dashed lines
correspond to the Julia-Zee dyon solution.}
\end{figure}

The two possibilities described above are idealised.
In reality, the transition  of  the  modified  dyon  into the Julia-Zee dyon is
accompanied by the emission of both vector and scalar bosons.
Calculations show that the modified  dyon  is stable to the transition into the
Julia-Zee dyon if the fraction of electric charge carried by the charged vector
bosons exceeds $0.33$ for $\kappa = 0$, and $0.24$ for $\kappa = 100$.

In Fig.~\ref{fig4}, we can see the ansatz  functions  corresponding  to the two
types of dyon solutions.
The most significant differences are observed for $j(\rho)$.
According to Eq.~(\ref{III:10}), the  large  $\rho$  asymptotics of this ansatz
function is determined by the electric charge of the dyon.
From  Fig.~\ref{fig4},  it   follows  that  for  a  given  $\tilde{\Omega}$ and
$\kappa$, the electric charge of the modified dyon is substantially larger than
that of the Julia-Zee dyon.

The ansatz function $u(\rho)$  is  also  different  for the two dyon solutions,
although not as much as $j(\rho)$.
Nevertheless, there is an important difference between  the two dyon solutions.
From  the  subplot  in  Fig.~\ref{fig4},  it  follows  that  for  the  modified
dyon, $u(\rho)$  crosses  zero  from above at finite $\rho$, reaches a minimum,
and then tends to zero from below.
At the same time, for  the  Julia-Zee  dyon,  $u(\rho)$  does not cross zero at
finite $\rho$, but tends monotonically to zero from above.
In contrast to  $j(\rho)$  and  $u(\rho)$,  the  ansatz  function  $h(\rho)$ is
practically the same for the two dyon solutions.

From Fig.~\ref{fig4}, we see that the maximum  of the ansatz function $f(\rho)$
is outside of the core of the Julia-Zee dyon.
Hence, the modified dyon consists of a core (deformed Julia-Zee dyon) surrounded
by a spherical cloud of the charged scalar condensate.
Note that in contrast  to  the  electrically charged Q-ball, the density of the
scalar condensate vanishes at the center of the  modified dyon, which is caused
by the boundary condition (\ref{III:4d}).

\section{Conclusion}                                              \label{seq:V}

We have shown  that  a   modified  dyon solution exists in  a non-Abelian gauge
model, which is essentially a modified Georgi-Glashow model.
The modification  consists  in  the  inclusion  of  a  self-interacting  scalar
isospinor field in the Lagrangian.
The sixth-order self-interaction potential of the scalar field is of the Q-ball
type.
This modification  of  the  Georgi-Glashow  model  leads  to  the  existence of
modified dyon solutions in addition to the usual Julia-Zee dyon solutions.
The modified dyons exist in  a  much narrower parametric domain compared to the
Julia-Zee dyons, and their properties differ significantly.

Firstly, except for the case $\kappa = 0$,  the  energy  and electric charge of
the modified dyons significantly exceed those of the Julia-Zee dyons.
Secondly, there is a non-zero minimum possible value of the electric charge for
the modified dyon, whereas this is equal to zero for the Julia-Zee dyon.
And thirdly,  both  the  minimum  and  maximum  value  of  the energy (electric
charge) of the modified dyon are achieved at the maximum value of the parameter
$\Omega$, whereas the energy (electric charge) of  the Julia-Zee dyon increases
monotonically with $\Omega$.
As for the solutions themselves, the greatest  differences are observed for the
ansatz  function  $j(\rho)$,  whose  derivative  determines  the electric field
strength of the dyon.

Finally, we note that various cosmological models tell us that the evolution of
the early Universe proceeded through a sequence of phase transitions.
Both  electrically charged magnetic  monopoles (dyons) and  clouds of a charged
scalar condensate can arise as a result of these phase transitions.
The interaction of a long-range  electric  field  of  the dyon with the charged
scalar condensate can lead to the formation of  modified dyons similar to those
considered in this paper.

\section*{Acknowledgements}

This work was supported by the Russian Science Foundation, grant No 23-11-00002.





\bibliographystyle{elsarticle-num}

\bibliography{article}






\end{document}